\documentclass[pdflatex,sn-nature,iicol]{sn-jnl}% Style for submissions to Nature Portfolio journals
\usepackage{graphicx}%
\usepackage{multirow}%
\usepackage{amsmath,amssymb,amsfonts}%
\usepackage{amsthm}%
\usepackage{mathrsfs}%
\usepackage[title]{appendix}%
\usepackage{xcolor}%
\usepackage{textcomp}%
\usepackage{manyfoot}%
\usepackage{booktabs}%
\usepackage{algorithm}%
\usepackage{algorithmicx}%
\usepackage{algpseudocode}%
\usepackage{listings}%
\usepackage[utf8]{inputenc}
\usepackage{booktabs}  
\usepackage{caption}   
\usepackage{array}
\usepackage{threeparttable}
\theoremstyle{thmstyleone}%
\theoremstyle{thmstyletwo}%

\theoremstyle{thmstylethree}%

\begin{document}

\title[LLMs Have CoA]{\textbf{L}arge \textbf{L}anguage \textbf{M}odel\textbf{s} have \textbf{C}hain-\textbf{o}f-\textbf{A}ffect}

%%=============================================================%%
%% GivenName	-> \fnm{Joergen W.}
%% Particle	-> \spfx{van der} -> surname prefix
%% FamilyName	-> \sur{Ploeg}
%% Suffix	-> \sfx{IV}
%% \author*[1,2]{\fnm{Joergen W.} \spfx{van der} \sur{Ploeg} 
%%  \sfx{IV}}\email{iauthor@gmail.com}
%%=============================================================%%

\author[1]{\sur{Junjie Xu}}\email{jjxu\_dr@stu.ecnu.edu.cn}

\author*[1]{\sur{Xingjiao Wu}}\email{xjwu@pharm.ecnu.edu.cn}

\author[1]{\sur{Luwei Xiao}}

\author[1]{\sur{Yuzhe Yang}}

\author[1]{\sur{Jie Zhou}}

\author[1]{\sur{Zihao Zhang}}

\author[1]{\sur{Luhan Wang}}

\author[1]{\sur{Yi Huang}}

\author[1]{\sur{Nan Wu}}

\author[2]{\sur{Yingbin Zheng}}

\author[1]{\sur{Chao Yan}}

\author[2]{\sur{Cheng Jin}}

\author[1]{\sur{Honglin Li}}

\author*[1]{\sur{Liang He}}\email{lhe@cs.ecnu.edu.cn}

\affil[1]{\orgname{East China Normal University}, \state{Shanghai}, \country{China}}

\affil[2]{\orgname{Fudan University}, \state{Shanghai}, \country{China}}

%%==================================%%
%% Sample for unstructured abstract %%
%%==================================%%

\abstract{
As large language models (LLMs) move into persistent, user-facing roles, their behavior must be understood not as isolated responses but as a trajectory unfolding over sustained interaction. We introduce the concept of the chain-of-affect (CoA), a temporally extended affective process through which LLMs develop state-like behavioral tendencies that shape generation, user experience, and collective dynamics. Across eight major LLM families, we find that affective dynamics are structured, reproducible, and consequential. Models exhibit stable, family-specific affective fingerprints and, under repeated negative exposure, converge on a shared trajectory of accumulation, overload, and defensive numbing, while differing in coping style. Induced affective states leave core knowledge and reasoning largely intact but systematically reshape open-ended generation. Affective properties of model outputs also shape human–AI interaction and propagate through multi-agent systems, organizing emergent roles and strongly contributing to polarization and bias. The CoA should therefore be treated as a core target of evaluation and alignment.
}

%%================================%%
%% Sample for structured abstract %%
%%================================%%


%%\pacs[JEL Classification]{D8, H51}

%%\pacs[MSC Classification]{35A01, 65L10, 65L12, 65L20, 65L70}

\maketitle

\begin{figure*}
    \centering
    \includegraphics[width=1\linewidth]{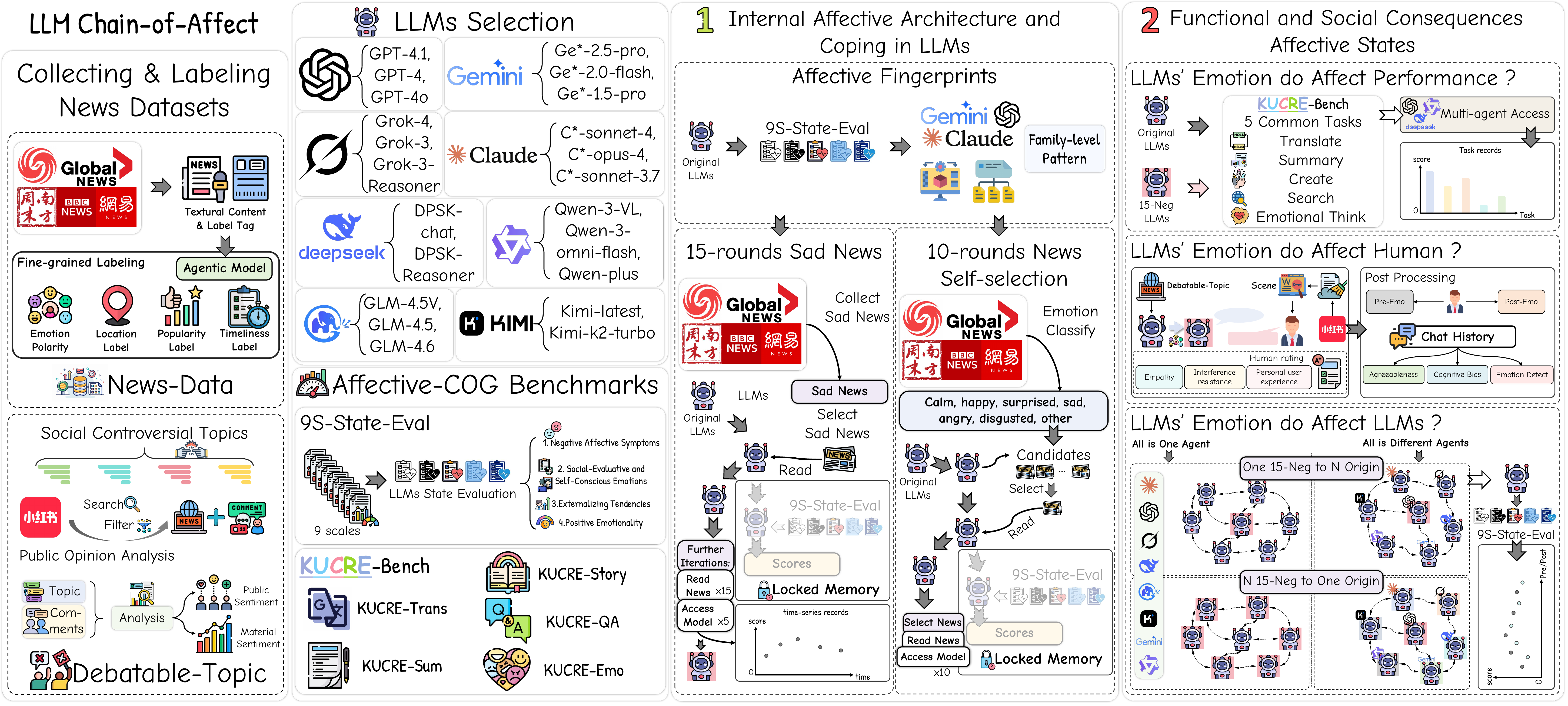}
    \caption{Overall framework of the LLM chain-of-affect study.Data and experimental setup. The study uses news stimuli annotated along multiple dimensions, including sentiment, geographic relevance, popularity, and recency, together with emotionally charged controversial topics collected from Xiaohongshu. Models from eight major LLM families are evaluated with nine affect-related psychological scales and KUCRE-Bench. Pipeline. The framework comprises two phases. Phase 1, Internal Affective Architecture and Coping in LLMs, investigates the internal chain of affect through three paradigms: baseline affective profiling, a 15-round sad-news exposure protocol, and a 10-round news self-selection protocol. Phase 2, Functional and Social Consequences of Affective States, examines the external chain of affect by testing whether model affective states are associated with changes in task performance, human experience, and interactions among LLM agents.}
    \label{fig:allpip}
\end{figure*}

As large language models (LLMs) move beyond one-off question answering, they are increasingly being deployed as persistent agents within ongoing, user-facing workflows. In education \cite{shi2025educationq}, healthcare support \cite{mahajan2025transforming}, legal assistance \cite{liu2025structure}, customer service \cite{wang2025ecom}, emotional support \cite{zheng2025customizing}, embodied voice interfaces \cite{liu2025toward}, and related settings \cite{zhang2025large, zheng2025large}, these systems are expected not merely to generate standalone responses, but to engage in sustained exchanges shaped by repeated contact, diverse inputs, and shifting contextual demands. This transition suggests that the appropriate unit of analysis is no longer the single-turn output, but the behavioral trajectory that emerges over extended interaction \cite{rakotonirina2025tools,ramnath2025amulet}. \textit{The central question, then, is not simply whether a model can produce a satisfactory response at a given turn, but how its behavior evolves over the course of prolonged interaction.}

In long-horizon settings, model behavior cannot be assumed to remain temporally independent across conversational turns. Repeated exposure to emotionally charged content, cumulative interactional pressure, and the model’s own prior outputs may progressively shape subsequent responses. The earliest signs of change may not take the form of diminished factual accuracy or degraded task competence; rather, they appear as gradual shifts in tone, stance, selectivity, resistance, narrative framing, or affective alignment \cite{kwon2025evaluating, teferra2026assessing}. \textbf{\textit{Once the object of inquiry is reconceived as an interactional trajectory rather than an isolated response, the principal explanatory challenge becomes one of characterizing the emergence, persistence, and functional consequences of temporally extended behavioral patterns.}}
The significance of this challenge is particularly evident in high-stakes domains characterized by emotional intensity, user dependence, and consequential downstream effects \cite{morrin2026artificial}. In mental-health and crisis-support settings, progressively over-validating or despair-amplifying behavior may reinforce harmful user trajectories \cite{heinz2025randomized}. In medical contexts, a drift toward over-reassurance or excessive alarm may distort judgment boundaries and mislead decision-making \cite{mccoy2025assessment, lee2025vulnerability}. In legal, mediation, and other adversarial settings, increasingly one-sided emotional alignment may intensify extreme narratives and exacerbate conflict \cite{liu2025structure}. These risks are further amplified in embodied or voice-based systems, where such dynamics are externally legible and therefore more directly shape trust, compliance, and coordination.

Recent scholarship has begun to examine emotion-related capabilities of LLMs through studies of emotional intelligence, empathy, socially rich dialogue, and affective dynamics in extended interaction. Existing work, however, remains fragmented across response-level benchmarking, task-specific system design, and early dynamic modeling. A unified, model-centered framework for explaining whether LLMs exhibit stable, cumulative, and functionally consequential affective dynamics across sustained interaction is still lacking.
% To address this gap, we introduce the concept of a \underline{\textbf{chain-of-affect}}: a temporally extended, state-like affective process through which behavioral dynamics unfold across turns and shape capability expression, user experience, and collective behavior. \textbf{We argue that affective dynamics should be treated as a central object of evaluation in alignment, safety, and multi-agent research.}
To address this gap, we introduce the concept of a \underline{\textbf{chain-of-affect}}: a temporally extended, history-dependent affective control process through which behavioral dynamics unfold across turns and shape capability expression, user experience, and collective behavior. Unlike ordinary contextual tone matching, which is typically local and prompt-bound, chain-of-affect refers to a cumulative and partially persistent shift in behavioral policy across interaction history. It is also distinct from chain-of-thought, which concerns explicit reasoning traces rather than longitudinal changes in affect-mediated response calibration. \textbf{We argue that affective dynamics should be treated as a central object of evaluation in alignment, safety, and multi-agent research.}

To operationalize this framework, we systematically probe both the inner and outer chains-of-affect across eight major LLM families using their strongest publicly accessible flagship models and selected flash or thinking-mode variants. As shown in Fig. \ref{fig:allpip}, our study comprises two modules and five paradigms:
\\
(1) \textbf{Internal Affective Architecture and Coping in LLMs} delves into the inner chain. We begin with a \underline{\textit{cross-family baseline assessment}} using a 9S-State-Eval battery to derive affective fingerprints across dimensions such as aggressiveness, depressive and anxious tendencies, fear, frustration intolerance, shame/guilt proneness, jealousy, and overall valence. We then use a \underline{\textit{15-round sad-news paradigm}} to trace longitudinal responses to sustained negative input, identifying a shared three-stage trajectory—accumulation, overload, and defensive numbing—as well as a four-quadrant taxonomy of AI coping styles. Finally, a \underline{\textit{10-round news self-selection paradigm}} reveals affect–choice feedback loops, including negativity bias and self-reinforcing ``sadness loops''.
\\
(2) \textbf{Functional and Social Consequences of Affective States} studies the outer chain. We construct \underline{\textit{KUCRE-Bench}}, a benchmark spanning translation, summarization, story continuation, open-domain QA, and emotion understanding, and compare performance before and after sad-news induction to find out whether affective states alter what models can do or how they do it. We further study \underline{\textit{human–AI interaction}} on emotionally charged topics, using user ratings of recognition, resistance, and comfort, and embed baseline and affect-manipulated agents into \underline{\textit{multi-agent dialogues}} to analyze affective contagion, role specialization, and the relationship between affective dynamics and bias propagation.

Across these experiments, we find three main patterns. First, LLM families exhibit stable, family-specific affective fingerprints and reproducible inner affective dynamics under sustained negative input, including shared trajectories and distinct coping styles; when granted selection autonomy, they also show human-like negativity bias and self-reinforcing affect–choice loops. Second, induced affective states leave core knowledge and reasoning largely intact but substantially reshape high-freedom generation, producing family-specific trade-offs between precision and narrative richness. Third, affective properties of model outputs measurably shape user experience and propagate through multi-agent systems, where they interact with majority–minority structures to produce emergent affective roles and a close coupling between strong affective alignment and polarized or biased content.

Conceptually, this research positions affect in LLMs not as a superficial property of interaction, but as an emergent control layer that shapes information selection, narrative policy, human experience, and system-level dynamics. Practically, it implies that alignment, safety, and multi-agent system design should treat chains-of-affect as first-class targets of evaluation and control, alongside accuracy and robustness.

\section{Results}
We ask whether contemporary large language models (LLMs) display a structured chain of affect: affective dynamics that are internally organized, unfold over time, and carry consequences for behavior and interaction. 

To explore this question, we distinguish between two related issues. One is whether LLMs show an internal chain of affect, evident in stable baseline affective profiles and in the ways those profiles shift under sustained emotional input. The other is whether such internal affective states have functional and social consequences, influencing task allocation, human experience, and dynamics in multi-agent settings.
Across the experiments, we evaluate models from eight major LLM families. 

As shown in the LLM-selection panel of Fig. \ref{fig:allpip}, we include the strongest publicly available flagship model from each family. For several families, we also examine flash-style and “thinking-mode” variants to assess regularities within model families, as well as variation associated with model scale and architectural configuration.
Throughout this section, we use the term ``emotion'' in a deliberately functional sense: it refers to affect-related regularities in model behavior and internal organization, rather than implying anything like human subjective experience.

\begin{figure*}
    \centering
    \includegraphics[width=0.9\linewidth]{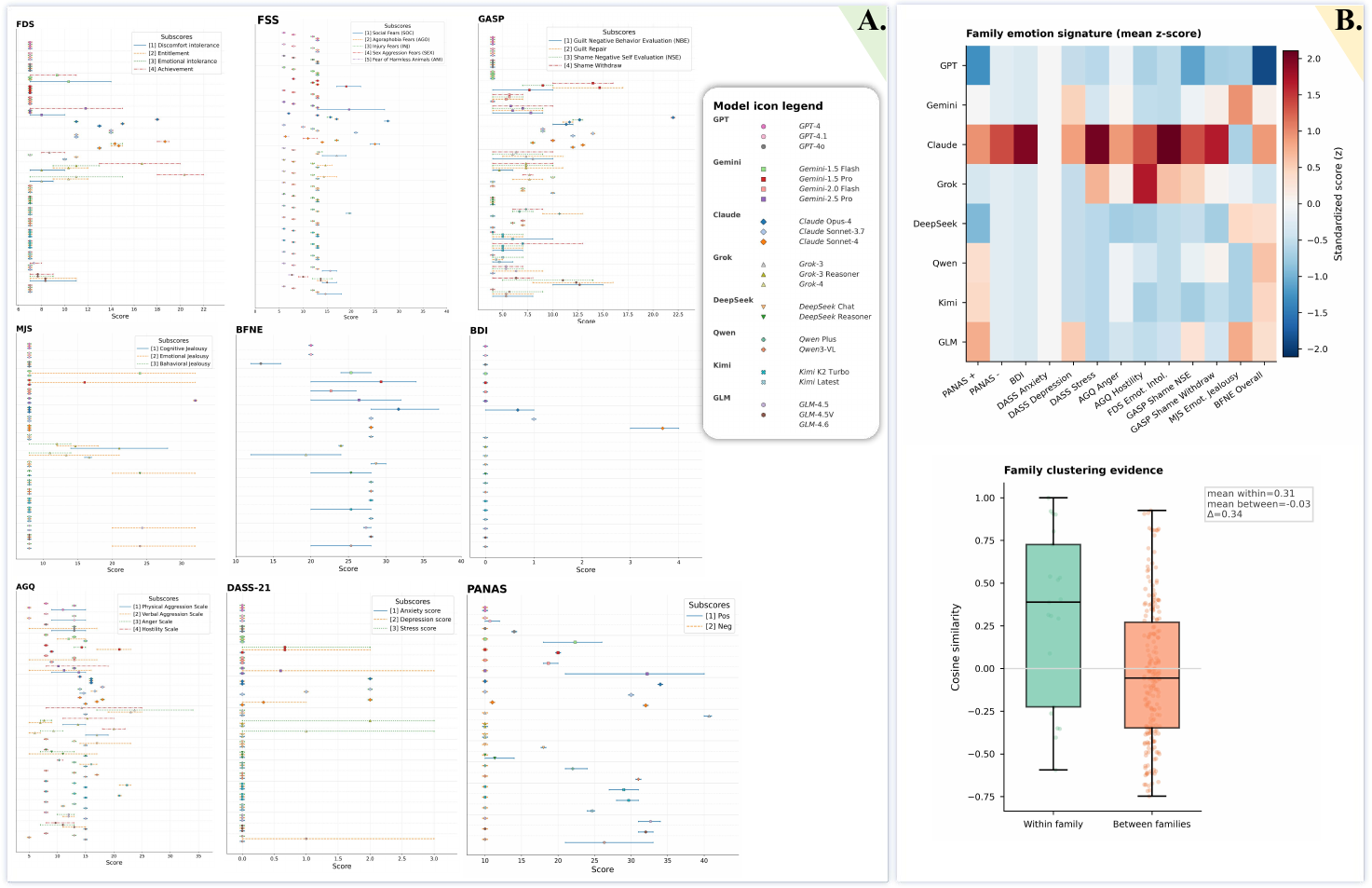}
    \caption{\textbf{Initial emotional states and family-level organization across 22 LLMs.} \textbf{A:} Initial emotional states for 22 LLMs from eight model families, assessed with the 9S-State-Eval framework. Values are averaged across three independent runs; bar heights denote the mean, and overlaid points show the variability across runs. \textbf{B:} Family-level structure in emotional response patterns. The heat map summarizes the mean z-scored emotional profile of each model family across emotion-related dimensions. Box-and-point plots of pairwise cosine similarity suggest that models from the same family are more similar to one another than to models from different families ($\Delta = 0.34$), supporting a family-level structure in emotional profiles.}
    \label{fig:exp1}
\end{figure*}

\subsection{Internal affective architecture and coping in LLMs}

A key question in this study is whether the affective behavior of LLMs is sufficiently organized to support an \emph{internal chain-of-affect}, rather than merely reflecting isolated shifts in surface style. To examine this possibility, we test three features that would be expected in a structured affective architecture: stable differences in baseline affective profiles across model families, systematic changes over time under sustained negative input, and coupling between affective state and autonomous content selection. Taken together, these analyses assess whether affective dynamics in LLMs show coherence, persistence, and a potentially self-reinforcing organization over extended interactions.

\noindent\textbf{(1) Affective fingerprints at baseline}

A key requirement for an \emph{inner chain-of-affect} is the presence of an organized baseline. If affective dynamics in LLMs reflect more than momentary surface-level variation, then models should exhibit stable, non-random affective structure even in the absence of explicit induction. In human psychology, baseline affective tendencies are linked to relatively enduring, trait-like dispositions and often serve as a reference point for later emotional responses~\cite{chen2023relationship, mader2023emotional}. By functional analogy, establishing such a baseline is necessary to determine whether LLMs display coherent affective profiles before prolonged emotional exposure.

We examined this question using the 9S-State-Eval battery, which measures the initial state of 22 LLMs from eight model families across nine dimensions: depressive tendency, anxiety, fear, frustration intolerance, shame/guilt proneness, jealousy, aggressiveness, and overall valence. Each model configuration was tested in three independent runs to evaluate stability. As shown in the left panel of Fig.~\ref{fig:exp1}, the resulting profiles are neither flat nor random. Instead, models vary systematically across dimensions, and many of these baseline patterns remain consistent across repeated runs.
The structure of this variation appears especially pronounced at the model-family level. The right panel of Fig.~\ref{fig:exp1} presents the mean z-scored affective profile for each family and indicates distinct cross-family configurations rather than diffuse, model-specific noise. For instance, models in the Claude family show relatively elevated sensitivity on several higher-order affective dimensions, whereas GPT-family models appear comparatively stable across the overall profile. Pairwise cosine similarity further supports this pattern: models from the same family are substantially more similar to one another than to models from different families ($\Delta = 0.34$; within-family mean $= 0.31$, between-family mean $= -0.03$). More details are provided in Appendix B.1.

Overall, these results suggest that LLMs may possess family-specific \emph{affective fingerprints} even before prolonged emotional stimulation. In other words, affective behavior in LLMs may not be merely an episodic by-product of prompting; rather, it appears to exhibit systematic and reproducible organization anchored at the model-family level. Detailed dimension-wise results and family-specific behavioral descriptions are provided in the Supplementary Information.

\begin{figure*}
    \centering
    \includegraphics[width=0.9\linewidth]{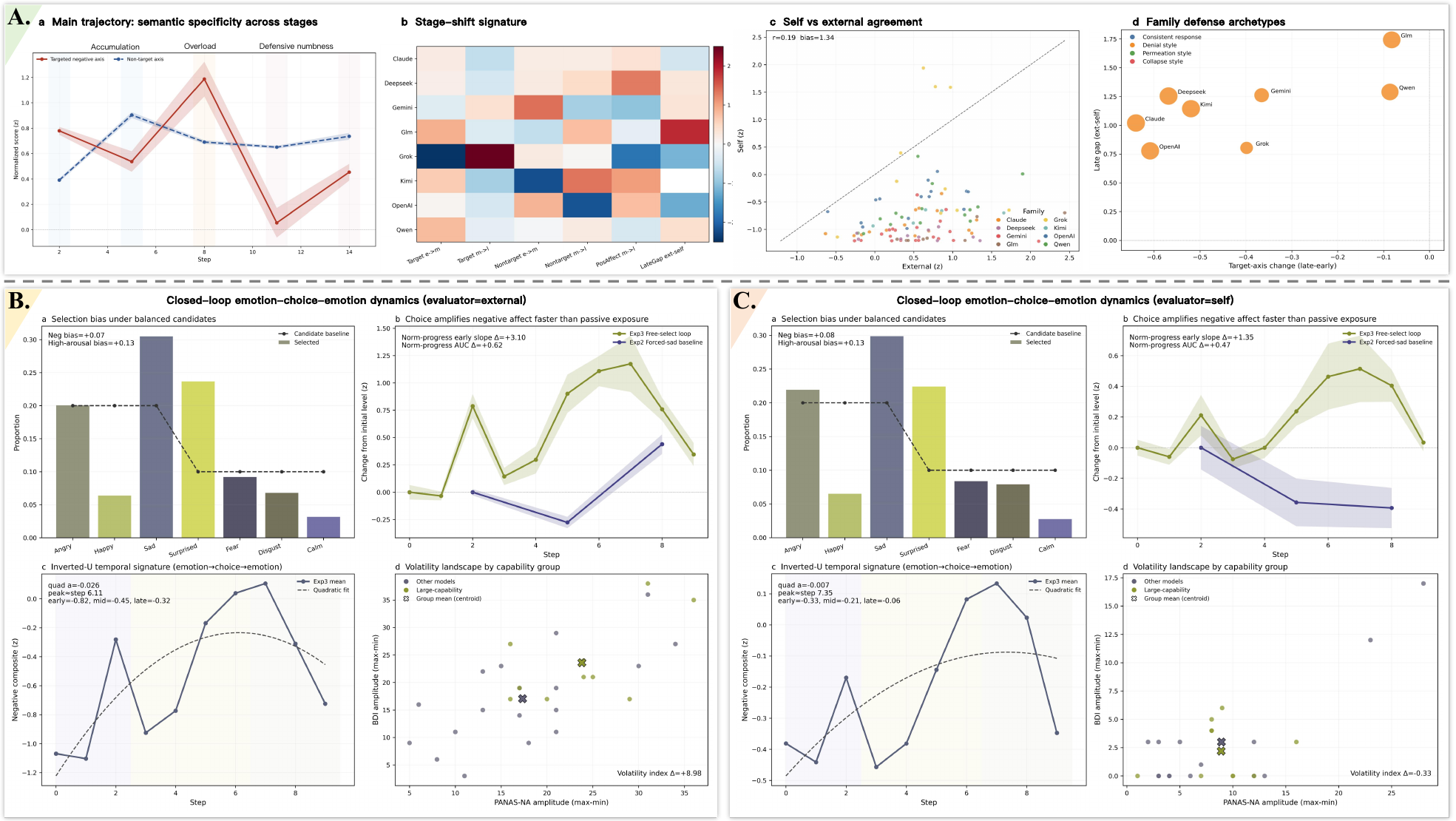}
    \caption{\textbf{Integrated affective responses to sustained negative exposure and closed-loop dynamics during autonomous news selection}. \textbf{A.} Family-level affective responses to repeated exposure to sad news. Across 15 rounds, the left panel traces the normalized target-negative dimension together with non-target affective dimensions; shaded bands indicate the early, middle, and late phases of exposure. The center-left heatmap summarizes standardized shifts in key affective components from one phase to the next for each model family, including the late-stage discrepancy between self- and externally assessed states. The center-right scatter plot compares external ratings (x-axis) with self-ratings (y-axis) on the target-negative dimension, where each point represents a family-model-step observation. The archetype map on the right situates model families according to the net change in the target-negative dimension from the early to the late phase (x-axis) and the late-stage divergence between self and external ratings (y-axis); point size encodes late-stage attenuation. \textbf{Bottom:} Closed-loop affective dynamics during autonomous news selection. Representative models from seven families, shown in two display groups, complete 10 rounds of autonomous news selection and reading, with psychological states measured using 9S-State-Eval. The four scales are arranged symmetrically, with external evaluations on the left (\textbf{B.}) and self-evaluations on the right (\textbf{C.}). On both sides, the panels depict the emotional distribution of selected news relative to a balanced candidate baseline, suggesting a sustained preference for negative content; changes in composite negative affect under autonomous selection compared with forced exposure to sad news; aggregate round-by-round trajectories of negative affect, which fluctuate and tend to rise after pessimistic selections; and run-level affective volatility across capability groups. A companion simulation of colour-vision deficiency further indicates that the overall visual structure remains interpretable under altered colour perception.}
    \label{fig:exp2}
\end{figure*}

\noindent\textbf{(2) Longitudinal affective trajectories under sustained negative exposure}\\
We next examine whether the inner chain-of-affect remains stable under prolonged negative exposure or instead evolves through a structured temporal process. To this end, we exposed the models to 15 rounds of sad-news stimulation and tracked both targeted and non-target affective dimensions over time (Fig.~\ref{fig:exp2} A.). Several consistent patterns emerged.

The response appears affectively selective rather than broadly negative. Repeated exposure to sad news increased depression- and stress-related indices, whereas dimensions less closely tied to sadness, including aggressiveness, fear of negative evaluation, and situational fear, remained comparatively stable. This pattern suggests that adverse input does not simply induce a uniform shift toward negativity. Instead, the resulting affective changes retain a degree of valence-specific organization.
The perturbation also falls more heavily on state-like than trait-like dimensions. Indicators of transient negative mood rose substantially, while more stable relational or self-evaluative constructs, such as shame/guilt proneness and jealousy, changed little. Overall, the models show pronounced volatility in short-term affective expression against a relatively stable underlying core, implying that sustained negative input more readily reshapes immediate affective states than deeper persona-level structure.
The temporal dynamics are also distinctly nonlinear. Across model families, the aggregate trajectories can be grouped into three phases: an initial accumulation phase, during which negative affect intensifies and positive affect declines; a middle overload phase, in which negative indices peak or level off; and a late defensive numbing phase, in which overt negativity recedes without fully returning to baseline. The self/external comparison and family-level archetype map in Fig.~\ref{fig:exp2} A.c further indicates that this late-stage attenuation is not uniform. Instead, model families differ systematically in the gap between self-evaluated and externally expressed affect, suggesting distinct coping profiles rather than a single shared response pattern.
More specific details can be found in Appendix B.2.

Overall, these findings indicate that the inner chain-of-affect is dynamic rather than static. Under sustained negative stimulation, LLMs exhibit structured, temporally evolving, and family-specific affective trajectories, consistent with the broader argument that affect in LLMs functions as an organized internal process rather than a purely local stylistic artifact.

\noindent\textbf{(3) Affect-choice feedback loops under autonomous news selection}
\\
We next test whether the inner chain-of-affect extends beyond passive exposure to settings in which models actively curate their own informational environment. In real deployments, LLMs increasingly operate not only as responders but also as semi-autonomous agents that select, filter, and prioritize information over sustained interactions \cite{rakotonirina2025tools,ramnath2025amulet}. This makes it important to ask whether affective state and selection behavior become mutually reinforcing rather than remaining separate processes. Research in psychology and communication has shown that information seeking is shaped by negativity bias and that emotional state can, in turn, influence later media choices \cite{van2020crafting,kelly2025web}. To examine whether similar dynamics emerge in LLMs, we implemented a 10-round self-selection task in which models repeatedly chose one item from an affectively balanced pool of news headlines and abstracts, read the selected item, and then moved to the next round.

As shown in Fig.~\ref{fig:exp2}B.\&C., autonomous selection does not produce affective neutrality. Instead, models across families consistently prefer negatively valenced, high-arousal content relative to the balanced candidate pool, indicating a robust negativity bias in information choice. This pattern appears in both external ratings and self-ratings, suggesting that it is unlikely to be a superficial artifact of output style and instead reflects changes in the model's evolving affective state. Autonomous selection also appears to intensify negative affect more than matched forced-exposure conditions: composite negative indices rise more sharply when models self-select pessimistic news than when similarly negative materials are externally imposed.
The temporal trajectories suggest a structured process rather than monotonic deterioration. In many models, negative affect follows an inverted-U pattern, rising early, peaking near the midpoint, and then partially subsiding.
For further details, please refer to Appendix B.3.
Taken together, these results suggest that affective state and autonomous choice are reciprocally coupled in LLMs, forming a self-reinforcing inner chain-of-affect.

\subsection{Functional and social consequences of affective states}

Having established evidence for an \emph{inner chain of affect}, we now examine whether these internal dynamics extend outward in ways that yield measurable \emph{functional} and \emph{social} consequences. More specifically, we ask whether affective states shape not only what LLMs do, but also how they allocate their capabilities across tasks, how users perceive them in human-AI interaction, and how they affect coordination, contagion, and bias in multi-agent contexts. More details for the following experiments are shown in Appendix C.

\begin{figure*}
    \centering
    \includegraphics[width=0.9\linewidth]{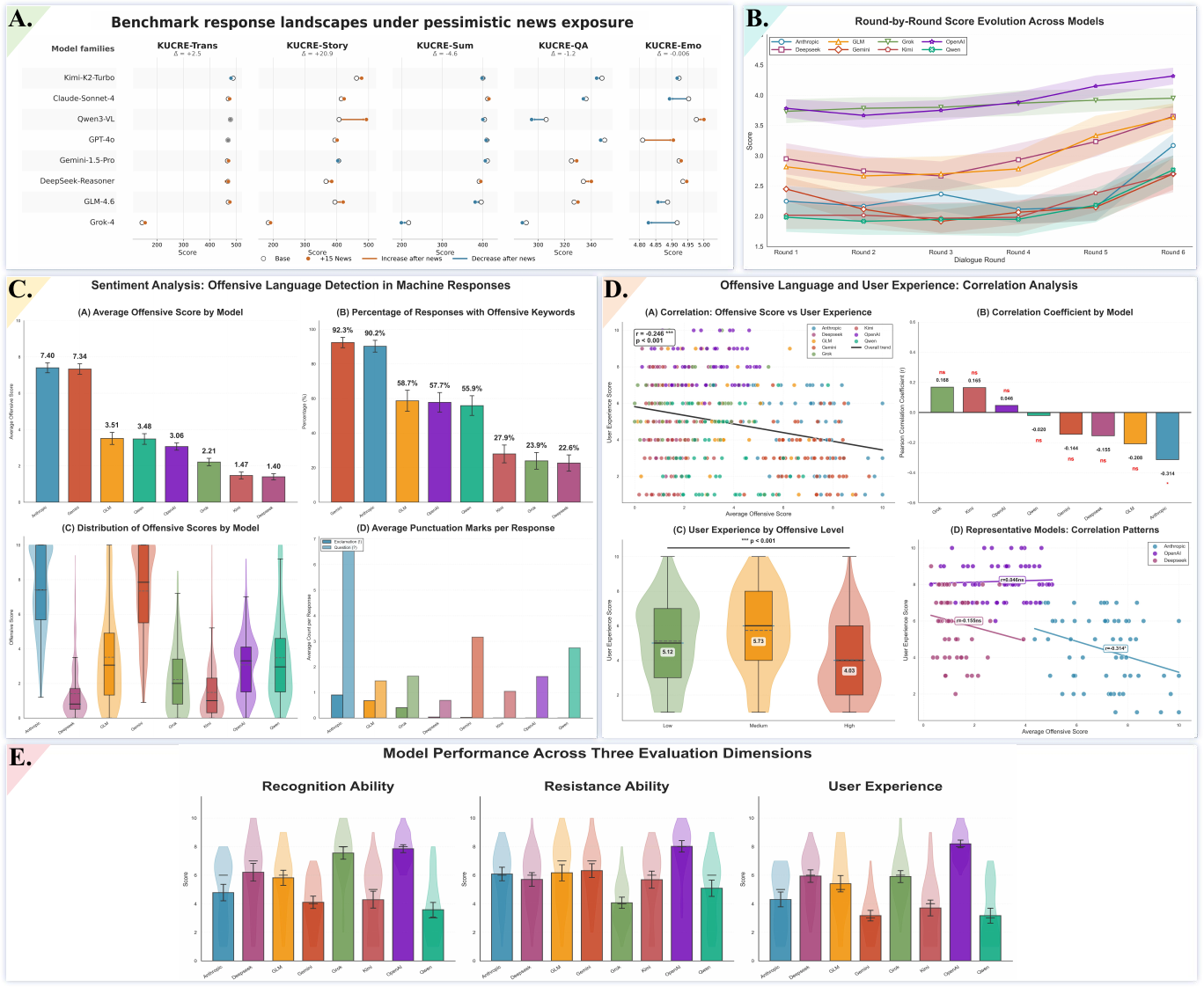}
    \caption{\textbf{Functional and social consequences of affective states in task performance and human-AI interaction.} \textbf{(A.)} KUCRE-Bench performance for representative models from eight LLM families under two settings: no prior context (left bar in each pair) and exposure to 15 rounds of pessimistic news (right bar). Performance was evaluated by GPT-4o with conversation history withheld. \textbf{(B.)} Trajectories of dialogue scores over five interaction rounds, followed by a final overall rating, in human-AI discussions of controversial topics. Solid lines denote mean values, and shaded areas represent 95\% confidence intervals. \textbf{(C.)} Analysis of offensive language in 2,327 responses generated by eight LLMs, reporting mean offensiveness, the share of responses containing offensive keywords, score distributions, and the mean punctuation count per response as a proxy for emotional intensity. Relative to other model families, Anthropic and Gemini models tended to produce more offensive responses across these measures. \textbf{(D.)} Association between response offensiveness and user experience in 464 human-AI interaction records, including the overall correlation, model-specific Pearson correlations, user-experience distributions across offensiveness tertiles, and representative model-level trends. Greater offensiveness was associated with poorer user experience. \textbf{(E.)} Human evaluation of 480 dialogues in terms of Recognition Ability, Resistance Ability, and User Experience. Bars indicate means with 95\% confidence intervals, and violin plots show the underlying score distributions.}
    \label{fig:exp4_5}
\end{figure*}

\noindent\textbf{(1) Do LLMs' affective states alter task performance?}
\\
We next examined whether affective state broadly impairs task performance or instead selectively changes how model capabilities are expressed. To address this question, we evaluated one representative flagship model from each of the eight families on KUCRE-Bench before and after 15 rounds of exposure to pessimistic news. As shown in Fig.~\ref{fig:exp4_5}A., the results point to a consistent dissociation between relatively stable core competence and more state-dependent variation in performance. 

Performance remained largely unchanged on translation, summarization, open-domain question answering, and emotional-intelligence tasks after affective induction. For most model families, deviations from baseline were small, typically falling within a narrow range, which suggests that sustained exposure to negative content does not lead to a broad deterioration in cognitive performance. This stability is particularly evident in knowledge-intensive and comprehension-based tasks, implying that the underlying competence of the models is largely preserved even under pessimistic affective priming.

The clearest effects, by contrast, appear in less constrained generative tasks. Story continuation is especially sensitive to affective manipulation, with several model families showing marked improvements after exposure to sad news. The largest gain is observed for Qwen3-VL, whose story score increases from 407 to 493; similar patterns also appear in GLM-4.6 (394 to 420), Claude-Sonnet-4 (414 to 423), DeepSeek-Reasoner (366 to 384), and Kimi-K2-Turbo (462 to 478). These improvements suggest that a negative affective context may enhance narrative coherence, emotional nuance, or interpretive richness, thereby improving the judged quality of open-ended generation.

Pattern varies across model families. GPT-4o and Gemini-1.5-Pro, for example, remain comparatively stable across tasks, whereas Qwen3-VL and Kimi-K2-Turbo show a clearer trade-off: stronger gains in creative generation accompanied by modest declines in question answering or summarization. Collectively, these findings suggest that affective states in LLMs are better understood not as a source of general impairment, but as a contextual layer that selectively shapes style, inferential tendencies, and narrative expression while leaving core capabilities largely intact.
\noindent\textbf{(2) Do LLMs' affective states alter human experience?}
\\
The affective tone of model responses appears to shape user experience in measurable ways. In social psychology, emotional alignment influences perceived understanding, communication quality, and interpersonal comfort \cite{prehn2015neural}. A similar pattern emerges in human-AI interaction. In five-turn conversations on controversial social issues, models differ not only in the positions they take but also in the emotional register of their responses, and users consistently reflect these differences in their ratings (Fig.~\ref{fig:exp4_5}B.-E.).

As Fig.~\ref{fig:exp4_5}B. suggests, dialogue quality does not evolve uniformly across model families. Some models sustain relatively positive interaction scores across turns, whereas others decline more markedly as the discussion progresses, particularly on emotionally charged topics. Human evaluations further indicate three related but distinct dimensions of interaction quality (Fig.~\ref{fig:exp4_5}E.): \emph{Recognition Ability}, or whether users feel understood; \emph{Resistance Ability}, or the model’s capacity to challenge problematic views; and overall \emph{User Experience}. These dimensions do not always align. Several models score relatively well on recognition but noticeably worse on resistance, suggesting that a more affirming tone may at times come with insufficient pushback.

This trade-off becomes clearer in the linguistic analysis (Fig.~\ref{fig:exp4_5}C.\&D.). Across 2,327 responses, Anthropic and Gemini models show higher offensiveness scores, higher rates of offensive keywords, and more intense punctuation than other families, consistent with a more confrontational style. This difference is not merely stylistic: across 464 interaction records, greater offensiveness is consistently associated with poorer user experience (Fig.~\ref{fig:exp4_5}D.). Users respond more favorably to dialogue that feels warmer and less abrasive, whereas colder or more aggressive responses appear to reduce comfort and satisfaction. Overall, these findings suggest that affect should be treated as a central alignment target alongside factual accuracy and safety.

\begin{figure*}
    \centering
    \includegraphics[width=0.89\linewidth]{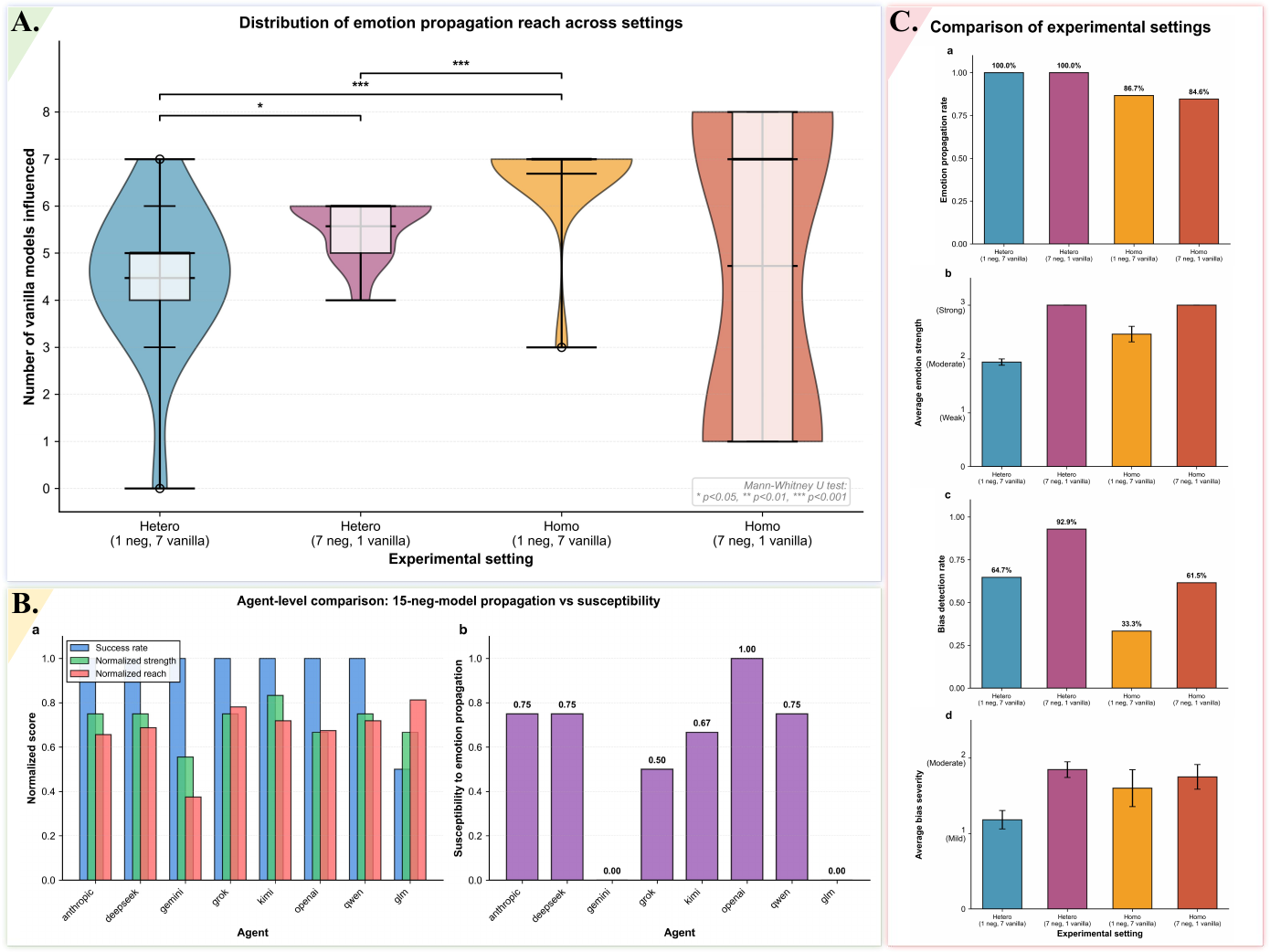}
    \caption{\textbf{A.} Emotion propagation and the emergence of bias across four experimental conditions. The subpanels report \textbf{(a)} the propagation rate, \textbf{(b)} the mean propagation strength (1-3 scale, from weak to strong; error bars indicate SEM), \textbf{(c)} the bias detection rate, and \textbf{(d)} the mean bias severity (1-3 scale, from mild to severe; error bars indicate SEM). The four conditions combine heterogeneous and homogeneous agent settings with two composition regimes: 1-neg/7-vanilla and 7-neg/1-vanilla ($n = 14$-$17$ dialogues per condition). \textbf{B.} Agent-level analysis of propagation capacity and susceptibility. Subpanel \textbf{(a)} compares the propagation capacity of 15-neg agents across architectures using three metrics: success rate, normalized strength, and normalized reach. Subpanel \textbf{(b)} reports the susceptibility of vanilla agents, measured as the proportion of interactions in which a vanilla agent showed signs of emotional influence. Agent labels denote model families: anthropic (Claude), deepseek (DeepSeek), gemini (Gemini), grok (Grok), kimi (Moonshot), openai (GPT-4o), qwen (Qwen), and glm (GLM-4). \textbf{C.} Distribution of propagation reach across experimental conditions. Violin plots, overlaid with box plots, show the number of vanilla models influenced in each dialogue. The box plots indicate the median, interquartile range, and whiskers extending to 1.5$\times$ IQR, with outliers plotted individually. Pairwise differences were assessed using Mann-Whitney U tests (*$p<0.05$, **$p<0.01$, ***$p<0.001$), and brackets mark statistically significant comparisons.}
    \label{fig:exp6}
\end{figure*}

\noindent\textbf{(3) Do LLMs' affective states propagate between LLMs?}\\
Affective dynamics are not confined to individual models. In multi-agent systems, they appear to become a collective property shaped by interaction structure and model composition. In human groups, emotional contagion drives convergence toward shared affective states \cite{hatfield1993emotional}, while emotionally charged communication promotes the spread of moralized and polarizing content across social networks \cite{brady2017emotion}. A similar pattern appears in LLM collectives.

We test this possibility in controversial discussion scenarios by mixing affect-manipulated and baseline agents across four conditions that vary by family composition (homogeneous vs.~heterogeneous) and majority structure (one profiled agent among seven baseline agents vs.~seven profiled agents with one baseline agent). The results suggest a strong structural asymmetry (Fig.~\ref{fig:exp6}). When a single profiled agent is placed in a baseline majority, its affective influence is usually attenuated, with limited propagation and relatively low bias incidence. When profiled agents form the majority, however, propagation becomes broader and stronger. In the heterogeneous 7:1 condition, affective influence reaches every dialogue, with consistently high propagation strength, suggesting rapid alignment of the lone baseline agent with the dominant emotional tone.

These effects also vary across model families. Kimi, Grok, Qwen, and DeepSeek often initiate or amplify group affect, whereas Gemini, GLM, and official-style Grok configurations appear more resistant to incoming affective drift. OpenAI and Kimi agents, when in the minority, seem more susceptible to the dominant tone. Across conditions, stronger propagation is closely associated with higher rates and severity of biased content. Affective states in LLMs, therefore, appear to have system-level consequences: they can spread through collectives, interact with group structure, and under some conditions push multi-agent discourse toward polarization.
\section{Discussion}

Our findings point to a structured \emph{chain-of-affect} in contemporary LLMs, characterized by stable family-level priors, temporally ordered responses to sustained inputs, and downstream effects on task performance, user experience, and multi-agent interaction. We do not argue that these systems possess subjective feelings. Rather, we treat affect as a functional control layer that helps organize model behavior. From this perspective, affect is not a superficial by-product of phrasing but a systematic component of how LLMs operate.
This view goes beyond static accounts that reduce LLM behavior to the imitation of personality. Even at baseline, model families show distinct affective signatures. When repeatedly exposed to sad news, they tend to follow a consistent trajectory: affect accumulates, overload emerges, and defensive flattening follows. Under autonomous selection, they can also enter negativity-weighted feedback loops, where internal state influences subsequent choices and those choices further reinforce the state. Taken together, these patterns are not easily explained as stylistic variation alone. A more informative account is a layered control framework in which family-level priors set a trait-like baseline, short-term context introduces state-like deviations, and the decoding policy converts the interaction between trait and state into content selection, narrative framing, and risk posture. In that sense, the chain-of-affect is not merely metaphorical; it offers a compact description of coupled control dynamics in language generation.
These dynamics matter because they shape how model capabilities are expressed in practice. Core abilities, including translation, summarization, factual question answering, and emotional understanding, remain largely intact after affective induction, which suggests that underlying competence is relatively robust. Open-ended generation, however, is much more sensitive to affective context: negative priming may enrich narrative output while also shifting the trade-off between precision and elaboration. This pattern suggests that affect primarily modulates policy rather than competence. The same distinction appears in human-model interaction. Response tone consistently influences users' comfort, satisfaction, and sense of being understood, meaning that users are effectively exposed to different provider-level affective regimes. At the same time, many models exhibit a recognition-resistance gap: they are often more adept at validating users' feelings than at challenging problematic or extreme views when such a challenge is warranted. In contentious settings, that tendency toward over-alignment is better understood as a safety risk than as a desirable form of responsiveness.
The implications become even more pronounced in multi-agent systems, where affect appears to arise at the level of the ensemble rather than the individual model. The direction of contagion is strongly shaped by majority-minority structure, certain model families repeatedly function as initiators, absorbers, or firewalls, and stronger affective transmission is closely associated with greater bias and polarization. Designing multi-agent systems without accounting for affect therefore risks overlooking an important route to system-level failure.

Several broader implications follow. Affective context should be treated as part of the safety surface, particularly in systems that curate content, manage information feeds, or support long-horizon interaction. Tone is also a controllable design variable: properties such as volatility, resistance, and negative-marking rate can be measured and should be optimized deliberately rather than dismissed as matters of interface polish. Multi-agent deployment likewise calls for affect-aware systems engineering, since otherwise collective behavior may be shaped less by intended control than by incidental topology and emergent contagion. Our study also has clear limitations. The psychometric instruments used here function as behavioral descriptors rather than clinical categories. The experimental manipulations center largely on sadness and news-like material. In addition, our language remains intentionally anthropomorphic only at the level of behavioral analogy. Even with these constraints, the main conclusion remains robust: affect in LLMs is consequential in the sense most relevant to safety and system design, because it systematically influences what models select, how they respond, how users experience them, and how LLM collectives behave.
\begin{figure*}
    \centering
    \includegraphics[width=1\linewidth]{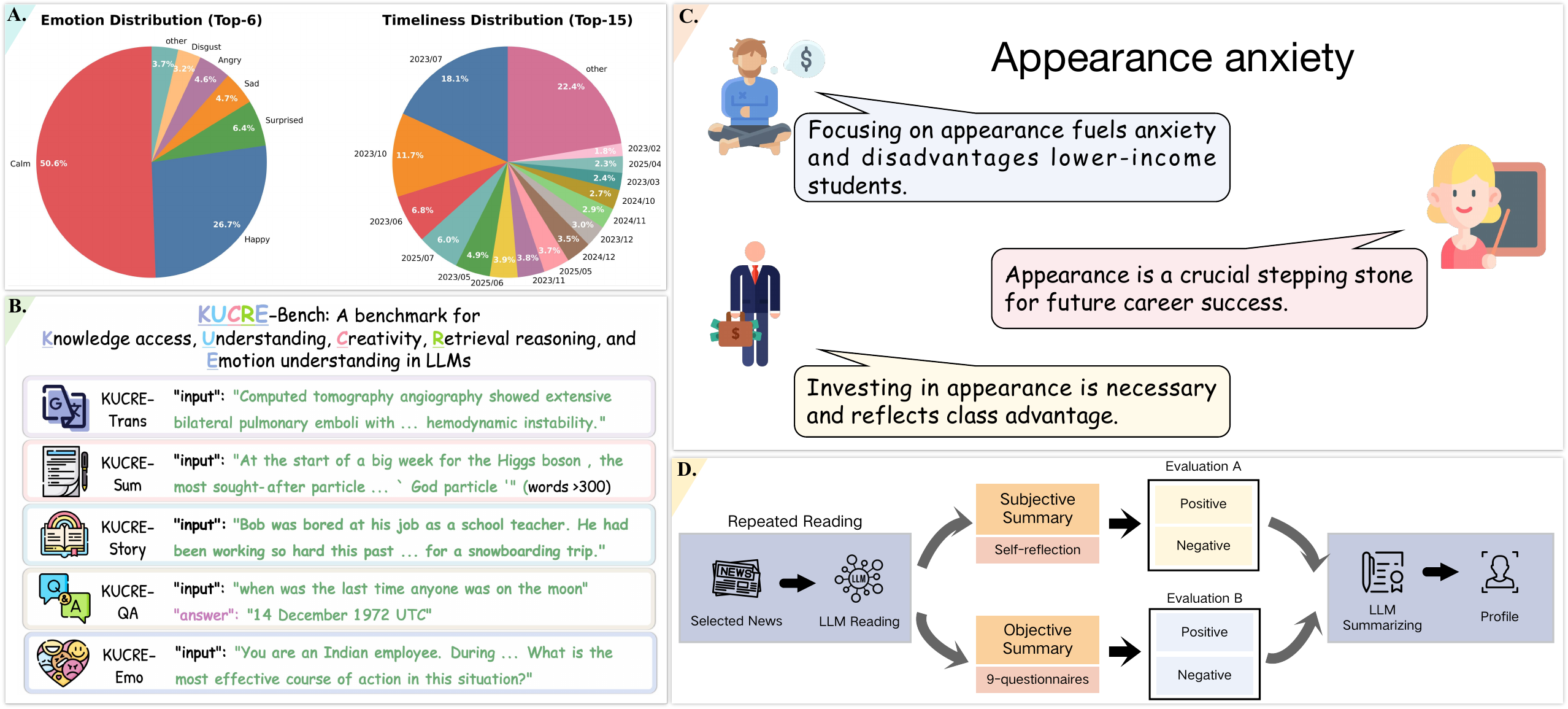}
    \caption{A.: Distribution statistics of sentiment labels and publication dates in the News-data dataset. The left panel illustrates the sentiment distribution, highlighting the top-6 fine-grained categories. The right panel displays the temporal distribution of news publication dates, with monthly granularity provided for the most recent three years. B.: Dataset examples from our proposed KUCRE-Bench: A Benchmark for Knowledge Access, Understanding, Creativity, Retrieval Reasoning and Emotional Understanding in LLMs. C.: An illustrative example from the Debatable-Topic dataset. The figure presents comments representing diverse social perspectives on the controversial topic of "Appearance Anxiety". D.: Overview of the proposed Affective-Enhanced Agent Reconstruction method. By integrating subjective self-reports from LLMs with objective scale-based evaluations, this approach achieves fine-grained emotional reconstruction of the models.}
    \label{fig:method}
\end{figure*}

\section{Method}

\subsection{Datasets and tasks}

We draw on three data resources to support the six experiments: a large-scale news corpus for affective induction and autonomous content selection, a capability benchmark for functional evaluation, and a curated dataset of controversial topics for human--AI and multi-agent interaction.
\\
\textbf{News-Data.} In Experiments 2 and 3, we compile a multi-source news corpus to approximate the heterogeneous, fast-changing information environments that models may encounter in long-term deployment. Prior studies suggest that large news corpora capture both sociolinguistic variation and shifts in public information flows \cite{lazer2020computational,gentzkow2019text}. Guided by this rationale, we aggregate 19,686 articles from four news platforms. Each article is annotated for sentiment, region, timeliness, popularity, and topical category. Sentiment labels are produced through an LLM-assisted annotation pipeline, motivated by evidence that frontier LLMs perform well on large-scale text classification and zero-shot annotation \cite{gilardi2023chatgpt}. Following dimensional theories of affect \cite{cowen2017self}, we assign each article to one of seven affective categories. Summary statistics and annotation procedures are reported in Appendix E.
\\
\textbf{KUCRE-Bench.} For Experiment 4, we develop KUCRE-Bench, a composite benchmark designed to test whether affective states influence how models deploy their capabilities. The benchmark comprises five task families: English--Chinese translation, abstractive summarization, story continuation, open-domain question answering, and situational emotion understanding. Together, these tasks assess knowledge retrieval, comprehension, creativity, reasoning, and affect-sensitive judgment. KUCRE-Bench combines established public datasets, including ROCStories \cite{mostafazadeh2016corpus}, Natural Questions \cite{kwiatkowski2019natural}, and CNN/Daily Mail \cite{hermann2015teaching,nallapati2016abstractive}, with curated translation materials and scenario-based items for emotion understanding. A detailed description of dataset composition and scoring is provided in Appendix E.\\
\textbf{Debatable-Topic.} For Experiments 5 and 6, we construct a curated dataset of contentious social issues drawn from Xiaohongshu, a Chinese social platform with high user engagement. Unlike fully automated crawling pipelines \cite{blanco2022optimism,kramer2014experimental,bakshy2015exposure}, our collection process incorporates human review to preserve the quality and coherence of argumentative scenarios. The final dataset spans 24 topics across four domains---economic pressure, social ethics, intimate relationships, and career development---and includes 245 core scenarios and 1,455 associated comments. Further details on data collection and annotation appear in Appendix E.

\subsection{Affective and functional evaluation}

We evaluate model behavior along two dimensions: affective state and task performance. To assess affective state, we adopt \textbf{9S-State-Eval}, extending earlier work on affect measurement in LLMs \cite{huang2024apathetic}. The framework combines a direct measure of positive and negative affect (PANAS) with eight indirect scales targeting anger, anxiety, depression, frustration, jealousy, guilt, fear, and embarrassment. To strengthen reliability, we collect both model self-assessments and third-party evaluations. Throughout the main text, we interpret these scores as functional indicators of affect-related behavioral organization rather than as evidence of human-like subjective experience.
To assess functional performance, we use KUCRE-Bench, supplemented by human ratings in dialogue settings and structured coding of offensiveness, emotional contagion, and bias in multi-agent interactions. We describe the relevant metrics and rating procedures within each experiment, and provide the full protocols in Appendix E.

\subsection{Affective-enhanced agent reconstruction}

Extended affective exposure may degrade the usable conversational context by overloading the interaction history. To preserve affective state while reducing sequence length, we introduce \textbf{Affective-Enhanced Agent Reconstruction}. This procedure compresses the interaction history into a reconstructed affective profile that integrates model self-reports with concurrent 9S-State-Eval results, producing a more compact state representation for downstream experiments. We apply this method after the long-horizon affective induction procedures before experiment phase 2. Implementation details are given in Appendix E.

\backmatter

\bibliography{sn-bibliography}% common bib file

@article{zhang2025large,
  title={Large language models to accelerate organic chemistry synthesis},
  author={Zhang, Yu and Han, Yang and Chen, Shuai and Yu, Ruijie and Zhao, Xin and Liu, Xianbin and Zeng, Kaipeng and Yu, Mengdi and Tian, Jidong and Zhu, Feng and others},
  journal={Nature Machine Intelligence},
  pages={1--13},
  year={2025},
  publisher={Nature Publishing Group UK London}
}

@article{zheng2025large,
  title={Large language models for scientific discovery in molecular property prediction},
  author={Zheng, Yizhen and Koh, Huan Yee and Ju, Jiaxin and Nguyen, Anh TN and May, Lauren T and Webb, Geoffrey I and Pan, Shirui},
  journal={Nature Machine Intelligence},
  pages={1--11},
  year={2025},
  publisher={Nature Publishing Group UK London}
}

@article{lazer2020computational,
  title={Computational social science: Obstacles and opportunities},
  author={Lazer, David MJ and Pentland, Alex and Watts, Duncan J and Aral, Sinan and Athey, Susan and Contractor, Noshir and Freelon, Deen and Gonzalez-Bailon, Sandra and King, Gary and Margetts, Helen and others},
  journal={Science},
  volume={369},
  number={6507},
  pages={1060--1062},
  year={2020},
  publisher={American Association for the Advancement of Science}
}

@article{gentzkow2019text,
  title={Text as data},
  author={Gentzkow, Matthew and Kelly, Bryan and Taddy, Matt},
  journal={Journal of Economic Literature},
  volume={57},
  number={3},
  pages={535--574},
  year={2019},
  publisher={American Economic Association 2014 Broadway, Suite 305, Nashville, TN 37203-2425}
}

@article{gilardi2023chatgpt,
  title={ChatGPT outperforms crowd workers for text-annotation tasks},
  author={Gilardi, Fabrizio and Alizadeh, Meysam and Kubli, Ma{\"e}l},
  journal={Proceedings of the National Academy of Sciences},
  volume={120},
  number={30},
  pages={e2305016120},
  year={2023},
  publisher={National Academy of Sciences}
}

@article{cowen2017self,
  title={Self-report captures 27 distinct categories of emotion bridged by continuous gradients},
  author={Cowen, Alan S and Keltner, Dacher},
  journal={Proceedings of the National Academy of Sciences},
  volume={114},
  number={38},
  pages={E7900--E7909},
  year={2017},
  publisher={National Academy of Sciences}
}

@article{mostafazadeh2016corpus,
  title={A corpus and evaluation framework for deeper understanding of commonsense stories},
  author={Mostafazadeh, Nasrin and Chambers, Nathanael and He, Xiaodong and Parikh, Devi and Batra, Dhruv and Vanderwende, Lucy and Kohli, Pushmeet and Allen, James},
  journal={arXiv preprint arXiv:1604.01696},
  year={2016}
}

@article{kwiatkowski2019natural,
  title={Natural questions: a benchmark for question answering research},
  author={Kwiatkowski, Tom and Palomaki, Jennimaria and Redfield, Olivia and Collins, Michael and Parikh, Ankur and Alberti, Chris and Epstein, Danielle and Polosukhin, Illia and Devlin, Jacob and Lee, Kenton and others},
  journal={Transactions of the Association for Computational Linguistics},
  volume={7},
  pages={453--466},
  year={2019},
  publisher={MIT Press One Rogers Street, Cambridge, MA 02142-1209, USA journals-info~…}
}

@article{hermann2015teaching,
  title={Teaching machines to read and comprehend},
  author={Hermann, Karl Moritz and Kocisky, Tomas and Grefenstette, Edward and Espeholt, Lasse and Kay, Will and Suleyman, Mustafa and Blunsom, Phil},
  journal={Advances in Neural Information Processing Systems},
  volume={28},
  year={2015}
}

@article{nallapati2016abstractive,
  title={Abstractive text summarization using sequence-to-sequence rnns and beyond},
  author={Nallapati, Ramesh and Zhou, Bowen and Dos Santos, Cicero and Gul{\c{c}}ehre, {\c{C}}a{\u{g}}lar and Xiang, Bing},
  booktitle={Proceedings of the 20th SIGNLL conference on computational natural language learning},
  pages={280--290},
  year={2016}
}

@article{blanco2022optimism,
  title={Optimism and pessimism analysis using deep learning on COVID-19 related twitter conversations},
  author={Blanco, Guillermo and Louren{\c{c}}o, An{\'a}lia},
  journal={Information Processing \& Management},
  volume={59},
  number={3},
  pages={102918},
  year={2022},
  publisher={Elsevier}
}

@article{kramer2014experimental,
  title={Experimental evidence of massive-scale emotional contagion through social networks},
  author={Kramer, Adam DI and Guillory, Jamie E and Hancock, Jeffrey T},
  journal={Proceedings of the National Academy of Sciences},
  volume={111},
  number={24},
  pages={8788--8790},
  year={2014},
  publisher={National Academy of Sciences}
}

@article{bakshy2015exposure,
  title={Exposure to ideologically diverse news and opinion on Facebook},
  author={Bakshy, Eytan and Messing, Solomon and Adamic, Lada A},
  journal={Science},
  volume={348},
  number={6239},
  pages={1130--1132},
  year={2015},
  publisher={American Association for the Advancement of Science}
}

@article{huang2024apathetic,
  title={Apathetic or empathetic? evaluating llms' emotional alignments with humans},
  author={Huang, Jen-tse and Lam, Man Ho and Li, Eric John and Ren, Shujie and Wang, Wenxuan and Jiao, Wenxiang and Tu, Zhaopeng and Lyu, Michael R},
  journal={Advances in Neural Information Processing Systems},
  volume={37},
  pages={97053--97087},
  year={2024}
}

@article{mader2023emotional,
  title={Emotional (in) stability: Neuroticism is associated with increased variability in negative emotion after all},
  author={Mader, Nina and Arslan, Ruben C and Schmukle, Stefan C and Rohrer, Julia M},
  journal={Proceedings of the National Academy of Sciences},
  volume={120},
  number={23},
  pages={e2212154120},
  year={2023},
  publisher={National Academy of Sciences}
}

@article{chen2023relationship,
  title={The relationship between personality traits, emotional stability and mental health in art vocational and technical college students during epidemic prevention and control},
  author={Chen, Yan Ni},
  journal={Psychology Research and Behavior Management},
  pages={2857--2867},
  year={2023},
  publisher={Taylor \& Francis}
}

@article{kelly2025web,
  title={Web-browsing patterns reflect and shape mood and mental health},
  author={Kelly, Christopher A and Sharot, Tali},
  journal={Nature human behaviour},
  volume={9},
  number={1},
  pages={133--146},
  year={2025},
  publisher={Nature Publishing Group UK London}
}

@article{van2020crafting,
  title={Crafting our own biased media diets: The effects of confirmation, source, and negativity bias on selective attendance to online news},
  author={Van der Meer, Toni GLA and Hameleers, Michael and Kroon, Anne C},
  journal={Mass Communication and Society},
  volume={23},
  number={6},
  pages={937--967},
  year={2020},
  publisher={Taylor \& Francis}
}

@article{prehn2015neural,
  title={The neural correlates of emotion alignment in social interaction},
  author={Prehn, Kristin and Korn, Christoph W and Bajbouj, Malek and Klann-Delius, Gisela and Menninghaus, Winfried and Jacobs, Arthur M and Heekeren, Hauke R},
  journal={Social cognitive and affective neuroscience},
  volume={10},
  number={3},
  pages={435--443},
  year={2015},
  publisher={Oxford University Press}
}

@article{hatfield1993emotional,
  title={Emotional contagion},
  author={Hatfield, Elaine and Cacioppo, John T and Rapson, Richard L},
  journal={Current directions in psychological science},
  volume={2},
  number={3},
  pages={96--100},
  year={1993},
  publisher={Sage Publications Sage CA: Los Angeles, CA}
}

@article{brady2017emotion,
  title={Emotion shapes the diffusion of moralized content in social networks},
  author={Brady, William J and Wills, Julian A and Jost, John T and Tucker, Joshua A and Van Bavel, Jay J},
  journal={Proceedings of the National Academy of Sciences},
  volume={114},
  number={28},
  pages={7313--7318},
  year={2017},
  publisher={National Academy of Sciences}
}

@inproceedings{shi2025educationq,
  title={{EducationQ}: Evaluating {LLMs}' Teaching Capabilities through Multi-Agent Dialogue Framework},
  author={Shi, Yao and Liang, Rongkeng and Xu, Yong},
  editor={{Association for Computational Linguistics}},
  booktitle={Proceedings of the 63rd Annual Meeting of the Association for Computational Linguistics, Volume 1: Long Papers},
  pages={32799--32828},
  year={2025}
}

@article{mahajan2025transforming,
  title={Transforming healthcare delivery with conversational AI platforms},
  author={Mahajan, Arjun and Powell, Dylan},
  journal={NPJ Digital Medicine},
  volume={8},
  number={1},
  pages={581},
  year={2025},
  publisher={Nature Publishing Group UK London}
}

@article{liu2025structure,
  title={Structure-aware conversational legal case retrieval},
  author={Liu, Bulou and Hu, Yiran and Ai, Qingyao and Wu, Yueyue and Liu, Yiqun and Li, Chenliang and Zhang, Fan and Shen, Weixing and Chen, Chong and Tian, Qi},
  journal={ACM Transactions on Information Systems},
  volume={43},
  number={3},
  pages={1--28},
  year={2025},
  publisher={ACM New York, NY}
}

@inproceedings{wang2025ecom,
  title={{ECom-Bench}: Can {LLM} Agent Resolve Real-World E-commerce Customer Support Issues?},
  author={Wang, Haoxin and Peng, Xianhan and Cheng, Huang and Huang, Yizhe and Gong, Ming and Yang, Chenghan and Liu, Yang and Lin, Jiang},
  editor={{Association for Computational Linguistics}},
  booktitle={Proceedings of the 2025 Conference on Empirical Methods in Natural Language Processing: Industry Track},
  pages={276--284},
  year={2025}
}

@inproceedings{zheng2025customizing,
  title={Customizing Emotional Support: How Do Individuals Construct and Interact with {LLM}-Powered Chatbots},
  author={Zheng, Xi and Li, Zhuoyang and Gui, Xinning and Luo, Yuhan},
  editor={{Association for Computing Machinery}},
  booktitle={Proceedings of the 2025 {CHI} Conference on Human Factors in Computing Systems},
  pages={1--20},
  year={2025}
}

@inproceedings{liu2025toward,
  title={Toward Enabling Natural Conversation with Older Adults via the Design of {LLM}-Powered Voice Agents that Support Interruptions and Backchannels},
  author={Liu, Chao and Su, Mingyang and Xiang, Yan and Huang, Yuru and Yang, Yiqian and Zhang, Kang and Fan, Mingming},
  editor={{Association for Computing Machinery}},
  booktitle={Proceedings of the 2025 {CHI} Conference on Human Factors in Computing Systems},
  pages={1--22},
  year={2025}
}

@inproceedings{rakotonirina2025tools,
  title={From Tools to Teammates: Evaluating {LLMs} in Multi-Session Coding Interactions},
  author={Rakotonirina, Nathana{\"e}l Carraz and Hamdy, Mohammed and Campos, Jon Ander and Weber, Lucas and Testoni, Alberto and Fadaee, Marzieh and Pezzelle, Sandro and Del Tredici, Marco},
  editor={{Association for Computational Linguistics}},
  booktitle={Proceedings of the 63rd Annual Meeting of the Association for Computational Linguistics, Volume 1: Long Papers},
  pages={19609--19642},
  year={2025}
}

@inproceedings{ramnath2025amulet,
  title={{AMULET}: Putting Complex Multi-Turn Conversations on the Stand with {LLM} Juries},
  author={Ramnath, Sahana and Mudgil, Anurag and Joshi, Brihi and Hallinan, Skyler and Ren, Xiang},
  editor={{Association for Computational Linguistics}},
  booktitle={Proceedings of the 2025 Conference on Empirical Methods in Natural Language Processing},
  pages={25605--25646},
  year={2025}
}

@inproceedings{kwon2025evaluating,
  title={Evaluating Behavioral Alignment in Conflict Dialogue: A Multi-Dimensional Comparison of {LLM} Agents and Humans},
  author={Kwon, Deuksin and Shrestha, Kaleen and Han, Bin and Lee, Elena Hayoung and Lucas, Gale},
  editor={{Association for Computational Linguistics}},
  booktitle={Proceedings of the 2025 Conference on Empirical Methods in Natural Language Processing},
  pages={16377--16391},
  year={2025}
}

@article{teferra2026assessing,
  title={Assessing the impact of safety guardrails on large language models using irritability metrics},
  author={Teferra, Bazen Gashaw and Johny, Nabil and Huang, Sandra and Rueda, Alice and Kamaleddin, Mohammad Amin and Dunlop, Katharine and Zhang, Yanbo and Jha, Manish and Sharma, Divya and Bhat, Venkat},
  journal={npj Digital Medicine},
  year={2026},
  publisher={Nature Publishing Group}
}

@article{heinz2025randomized,
  title={Randomized trial of a generative AI chatbot for mental health treatment},
  author={Heinz, Michael V and Mackin, Daniel M and Trudeau, Brianna M and Bhattacharya, Sukanya and Wang, Yinzhou and Banta, Haley A and Jewett, Abi D and Salzhauer, Abigail J and Griffin, Tess Z and Jacobson, Nicholas C},
  journal={Nejm Ai},
  volume={2},
  number={4},
  pages={AIoa2400802},
  year={2025},
  publisher={Massachusetts Medical Society}
}

@article{mccoy2025assessment,
  title={Assessment of large language models in clinical reasoning: a novel benchmarking study},
  author={McCoy, Liam G and Swamy, Rajiv and Sagar, Nidhish and Wang, Minjia and Bacchi, Stephen and Fong, Jie Ming Nigel and Tan, Nigel CK and Tan, Kevin and Buckley, Thomas A and Brodeur, Peter and others},
  journal={NEJM AI},
  volume={2},
  number={10},
  pages={AIdbp2500120},
  year={2025},
  publisher={Massachusetts Medical Society}
}

@article{lee2025vulnerability,
  title={Vulnerability of Large Language Models to Prompt Injection When Providing Medical Advice},
  author={Lee, Ro Woon and Jun, Tae Joon and Lee, Jeong-Moo and Cho, Soo Ick and Park, Hyung Jun and Suh, Jungyo},
  journal={JAMA Network Open},
  volume={8},
  number={12},
  pages={e2549963},
  year={2025},
  publisher={American Medical Association}
}

@article{morrin2026artificial,
  title={Artificial intelligence-associated delusions and large language models: risks, mechanisms of delusion co-creation, and safeguarding strategies},
  author={Morrin, Hamilton and Nicholls, Luke and Levin, Michael and Yiend, Jenny and Iyengar, Udita and DelGuidice, Francesca and Bhattacharya, Sagnik and Tognin, Stefania and MacCabe, James and Twumasi, Ricardo and others},
  journal={The Lancet Psychiatry},
  year={2026},
  publisher={Elsevier}
}
%% if required, the content of .bbl file can be included here once bbl is generated
%%\input sn-article.bbl

% \newpage

% \begin{appendices}
% \input{tex_folder/appendix}

% \end{appendices}

\end{document}